\documentstyle[11pt]{article}

\newcommand{\beq}{\begin{equation}}
\newcommand{\eeq}{\end{equation}}
\newcommand{\beqa}{\begin{eqnarray}}
\newcommand{\eeqa}{\end{eqnarray}}
\newcommand{\inte}{\int_{-\infty}^{\infty}d}

\begin{document}
\baselineskip14pt
\begin{flushright}ITP-SB-92-60\end{flushright}
\begin{center}
{\Large\bf Thermodynamics of the 3-State Potts Spin Chain}

\vskip.3in
{\large Rinat Kedem\footnote{e-mail: rinat@max.physics.sunysb.edu}}\\
Institute for Theoretical Physics, State University of New York\\
Stony Brook, NY 11794
\end{center}

\vskip.5in

\begin{abstract}
We demonstrate the relation of the infrared anomaly of conformal field
theory with entropy considerations of finite temperature
thermodynamics for the 3-state Potts chain.  We compute the free energy
and compute the low temperature specific heat for both the
ferromagnetic and anti-ferromagnetic spin chains, and find the central
charges for both.
\end{abstract}

\section{Introduction}

The conformal field theory treatment of quantum spin chains at
positive temperature $T$ deals with systems of size $M$ in the limit
\beq
 M\rightarrow\infty, \qquad T\rightarrow 0,\qquad  T M\ {\rm fixed}.
\label{cft limit}
\eeq
This limit is discussed in terms of a variable $q=\exp(2 \pi v/ M T)$
where $v$ is the speed of sound. The modular invariant partition
function is computed in terms of $q$, and one of the important
results~\cite{bcn,affleck} is that as $q\rightarrow 1$, the free energy
per site is given as
\beq
f=e_{GS}-{c\pi\over 6v}T^2  + o(T^2)
\eeq
where $c$ is the central charge as determined from finite size
corrections to the ground state energy
\beq
E_{GS}=Me_0-{\pi c v\over 6 M} + o(M^2).
\eeq

However the limit~(\ref{cft limit}) is not the limit of
thermodynamics. This limit is
\beq
 M\rightarrow\infty, \qquad T\ {\rm fixed}.
\label{tba limit}
\eeq
Here the low temperature behavior of the specific heat is obtained by
letting $T\rightarrow 0$ after the limit~(\ref{tba limit}) is taken.

These two limits will give the same result if there are no additional
length scales in the problem. In this case the result is
obtained~\cite{bcn,affleck} that the specific heat $C$ is
\beq
C\sim \frac{\pi c}{3 v} T.
\label{spheat}
\eeq

For the anti-ferromagnetic 3-state Potts chain, the low lying order
one excitations in the limit~(\ref{cft limit}), were used in~\cite{km}
to compute the partition function, and the
result~(\ref{spheat}) was obtained.  This was accomplished by
adding up the order one excitations found from the Bethe's
equations for the model~\cite{albert,adm2}.  The partition function
obtained in this manner is the modular invariant partition function of
conformal field theory.

The counting of states in~\cite{km} depends on the fact that the
momenta of order one excitations obey not only a fermi exclusion rule
$P_j\neq P_k$, but also have additional exclusion rules: The number of
states near $e(P)=0$ diminishes as the number of order one excitations
in the system increases.  Correct counting of the states,
incorporating these exclusion rules, gives a partition function which
has a central charge smaller than that of fermions.  In the case of
the anti-ferromagnetic 3-state Potts model, there are three kinds of
quasi-particle excitations, which, for purely fermionic exclusion
rules, would give a central charge of 3/2, whereas the actual central
charge of the model is 1.

In this paper, we use the thermodynamic limit~(\ref{tba limit}) to
obtain the low temperature specific heat of the 3-state Potts chain
from Bethe's equations, using the methods
of~\cite{yny,gaudin,takahashi}.  In section 2, we write the Bethe
equations for the finite lattice and introduce the completeness
rules~\cite{adm} for $Q=0$.  We find it convenient to study the
ferromagnetic and antiferromagnetic cases using two different sets of
integral equations.  In section 3, we write the free energy for the
ferromagnetic case in terms of only one integral equation, and compute
the linear term in the low temperature specific heat.  In section 4,
we do the same for the anti-ferromagnetic chain, this time in terms of
two integral equations.  We obtain the central charge of the conformal
limit of both spin chains from the linear term in the specific heat,
using equation~(\ref{spheat}).  In section 5, we consider the sector
$Q=1$.

In section 6, we discuss how the  counting of states in the
finite size system~\cite{km} is incorporated into the discussion in
terms of densities in the thermodynamic limit.  We find that in the
thermodynamic limit, the counting of states is contained in the
entropy, written in terms of densities.  The densities are related
through the thermodynamic limit of the Bethe
equations~(\ref{bae+}),~(\ref{bae-}), and this relationship corresponds
to the way the number of available states depends on the number of
excitations in the finite size system. These equations become
particularly important when evaluating the low temperature specific
heat, where  we need to consider the equations in precisely the
limit which corresponds to $P\rightarrow0$ in order to extract the
linear term in the specific heat.

It is important to note that the 3-state Potts model is the $D_4$
model in the classification of Pasquier~\cite{pasquier}, which, by
orbifold construction~\cite{fg,roche}, has certain sectors of eigenvalues
which overlap with the $A_5$ RSOS model.
{}From this construction, it is to be expected that the thermodynamic
quantities of the two models are equal as long as the sectors which
dominate the thermodynamics are common to the sectors which overlap.
Indeed, the ground states of both the ferromagnetic and
anti-ferromagnetic ends of the $D_4$ model are the same as the
critical $A_5$ model at the boundaries of the $III/IV$ and $I/II$
regimes respectively. Thus the central charges computed for
$A_5$~\cite{abf,huse,baz-resh} coincide with the central charges of
the $D_4$ model~\cite{fl,pearceun}.  We further note that the
classical two dimensional anti-ferromagnetic 3-state Potts model which
is critical at $T=0$ is equivalent~\cite{baxter,saluer,park-widom} to
the 3-coloring problem and also has central charge
$c=1$~\cite{kim-pearce}.

  The thermodynamics of the $A_n$ series was studied
in~\cite{baz-resh}, where the integral equations for the free energy,
the central charges and order one excitations above the ground state
were found. The thermodynamics of other affine Lie algebras were also
studied in~\cite{kuniba}.  In this paper, however, we obtain  different sets of
integral equations, which display a more direct relationship to the
modular invariant partition function discussed in~\cite{km}
for the anti-ferromagnetic chain and in~\cite{dkm} for the ferromagnetic chain.
  The fact
that there are different sets of integral equations for the model is
related to the fact that the modular invariant partition function can
be expressed as sums over different sets of quasi-particle
excitations~\cite{kkmm}.

\section{Formulation}
\setcounter{equation}{0}
The three state Potts hamiltonian is
\beq
H = \pm\frac{2}{\sqrt{3}} \sum_{j=1}^{M}\bigl\{ X_j + X_j^{\dagger}
+ Z_j Z_{j+1}^{\dagger} + Z_j^\dagger Z_{j+1}\bigr\}
\label{ham}
\eeq
with periodic boundary conditions, where $M$ is the number of sites in
the chain, and the matrices $X_j$ and $Z_j$ are
\beqa
X_j &=&I \otimes I \otimes \ldots \otimes  X_{j^{th}} \otimes \ldots\otimes I,
\nonumber\\
Z_j &=&I \otimes I \otimes \ldots \otimes  Z_{j^{th}} \otimes \ldots\otimes I,
\eeqa
$I$ being the $3 \times 3$ identity matrix and $X$ and $Z$ are $3\times3$
matrices with entries:
\beq
X_{ij} = \delta_{i,j+1}(\hbox{\rm mod}\ 3),\qquad
Z_{ij} = \delta_{i,j} \omega^{i-1},\qquad\omega=e^{2 \pi i/3}.
\eeq

The hamiltonian with the ($+$) $-$ sign is referred to as the
(anti-)ferromagnetic Potts chain
 It
commutes with the spin rotation operator, whose eigenvalue is $e^{2\pi
i Q/3}$, $Q=0,\pm1$.  The eigenvalues of the hamiltonian~(\ref{ham}) are
derived from functional equations~\cite{abf,baz-resh,bazstrog,bbp,amp} from
which we find~\cite{albert}:
\beq
E=\sum_{j=1}^{L} \cot{(i \lambda_j + \frac{\pi}{12})} - \frac{2 M}{\sqrt3},
\qquad L=2(M-|Q|),\quad Q=0,1,2
\label{energy}
\eeq
where the set $\{\lambda_j\}$  satisfy the Bethe equations:
\begin{equation}
\label{bae1}
\Bigl[\frac{\sinh(i \pi/12-\lambda_j)}{\sinh(i\pi/12+\lambda_j)}
\Bigr]^{2M}
=(-1)^{M+1}  \prod_{k=1}^L \frac{\sinh(i \pi/3-(\lambda_j-\lambda_k))}
{\sinh(i \pi/3+(\lambda_j-\lambda_k))},\ \ j=1,\ldots,L
\end{equation}

Not all solutions of the equations~(\ref{bae1}) correspond to
eigenvalues of the hamiltonian~(\ref{ham}).  The equations do not
impose sufficient restrictions on the set $\{\lambda_j\}$.  There are
additional conditions, which ensure that the energy is real,
postulated from finite size studies in~\cite{adm}, where the spectrum
of $H$ was classified.  We introduce these conditions by
writing~(\ref{bae1}) in logarithmic form.  The solutions
of~(\ref{bae1}) which correspond to eigenstates of the
hamiltonian~(\ref{ham}) fall into 5 classes, where below $\lambda_j$
is a real number:
\beq
\lambda_j^+  =  \lambda_j,\quad
\lambda_j^-  =  \lambda_j + \frac{i\pi}{2},\quad
\lambda_j^{2s} = \lambda_j \pm \frac{i\pi}{6},\quad
\lambda_j^{-2s} = \lambda_j \pm \frac{i\pi}{3},\quad
\lambda_j^{ns} = \lambda_j \pm \frac{i\pi}{4}.
\label{roottypes}
\eeq
The last three always occur in complex conjugate pairs.  Since we are
interested in the infinite lattice limit, the imaginary parts are
assumed to be exact.  We denote the number of each type of root
$\alpha\in\{+,-,2s,-2s,ns\}$ by $m_\alpha$.

We rewrite the Bethe equations~(\ref{bae1}) to explicitly display the
different types of roots~(\ref{roottypes}).  Let
\beq
h(\lambda) = \frac{\sinh(\frac{i\pi}{3}-\lambda)}
                  {\sinh(\frac{i\pi}{3}+\lambda)}.
\eeq
Then~(\ref{bae1}) become, for $\alpha=+$ or $-$:
\beqa
(-1)^{M+1}\Bigl[\frac{\sinh(\frac{i\pi}{12}-\lambda_j^\alpha)}
           {\sinh(\frac{i\pi}{12}+\lambda_j^\alpha)}\Bigr]^{2M} & = &
\prod_{k=1}^{m_+}h(\lambda_j^\alpha-\lambda_j^+)
\prod_{k=1}^{m_-}h(\lambda_j^\alpha-\lambda_j^-)
\prod_{k=1}^{m_{2s}}
h(\lambda_j^\alpha-\lambda_j^{2s})h(\lambda_j^\alpha-\lambda_j^{2s*})
\nonumber \\ & &\hbox{\hskip-1in}\times
\prod_{k=1}^{m_{-2s}}
h(\lambda_j^\alpha-\lambda_j^{-2s})h(\lambda_j^\alpha-\lambda_j^{-2s*})
\prod_{k=1}^{m_{ns}}
h(\lambda_j^\alpha-\lambda_j^{ns})h(\lambda_j^\alpha-\lambda_j^{ns*})
\label{bae2.1}
\eeqa
whereas for $\beta=2s,-2s$ or $ns$, the equations for each complex conjugate
pair are multiplied together:
\beqa
\Bigl[\frac{\sinh(\frac{i\pi}{12}-\lambda_j^\alpha)
            \sinh(\frac{i\pi}{12}-\lambda_j^{\alpha*})}
           {\sinh(\frac{i\pi}{12}+\lambda_j^\alpha)
            \sinh(\frac{i\pi}{12}+\lambda_j^{\alpha*})}\Bigr]^{2M} &=&
\prod_{k=1}^{m_+}h(\lambda_j^{\alpha}-\lambda_k^+)h(\lambda_j^{\alpha*}-\lambda_k^+)
\prod_{k=1}^{m_-}h(\lambda_j^{\alpha}-\lambda_k^-)h(\lambda_j^{\alpha*}-\lambda_k^-)
\nonumber \\ & &\hbox{\hskip-1in}\times
\prod_{k=1}^{m_{2s}}h(\lambda_j^{\alpha}-\lambda_k^{2s})
                    h(\lambda_j^{\alpha*}-\lambda_k^{2s})
                    h(\lambda_j^{\alpha}-\lambda_k^{2s*})
                    h(\lambda_j^{\alpha*}-\lambda_k^{2s*})
\nonumber \\ & &\hbox{\hskip-1in}\times
\prod_{k=1}^{m_{-2s}}h(\lambda_j^{\alpha}-\lambda_k^{-2s})
                    h(\lambda_j^{\alpha*}-\lambda_k^{-2s})
                    h(\lambda_j^{\alpha}-\lambda_k^{-2s*})
                    h(\lambda_j^{\alpha*}-\lambda_k^{-2s*})
\nonumber \\ & &\hbox{\hskip-1in}\times
\prod_{k=1}^{m_{ns}}h(\lambda_j^{\alpha}-\lambda_k^{ns})
                    h(\lambda_j^{\alpha*}-\lambda_k^{ns})
                    h(\lambda_j^{\alpha}-\lambda_k^{ns*})
                    h(\lambda_j^{\alpha*}-\lambda_k^{ns*}).
\label{bae2.2}
\eeqa

We follow ref.~\cite{adm} in taking the logarithm of
equations~(\ref{bae2.1}),~(\ref{bae2.2}).  To do this, we define the
functions $t_\alpha$ and $\Theta_{\alpha,\beta}$ in the following way:
\beq
t_\pm(\lambda_j^\pm) =\cases{
 -2i \ln\Bigl[\pm\frac{\sinh(i\pi/12-\lambda_j^\pm)}
                    {\sinh(i\pi/12+\lambda_j^\pm)}\Bigr] & for $\alpha=\pm$;\cr
  - 2i f_\alpha\ln\Bigl[\frac{\sinh(i\pi/12-\lambda_j^\alpha)}
                   {\sinh(i\pi/12+\lambda_j^\alpha)}
              \frac{\sinh(i\pi/12-\lambda_j^{\alpha*})}
                   {\sinh(i\pi/12+\lambda_j^{\alpha*})}\Bigr]&
for $\alpha=\pm 2s,ns $\cr}
\eeq
where $f_{\pm2s}=-1$  $f_{ns}=-1/2$ and $f_\pm=1$. For  $\alpha,\beta=\pm$:
\beq
\label{theta}
\Theta_{\alpha\beta}(\lambda_j^\alpha-\lambda_k^\beta)=
-i \ln\Bigl[\alpha\beta h(\lambda_j^\alpha-\lambda_k^\beta)\Bigr]
\eeq
For $\alpha=\pm$, $\beta=\pm 2s, ns$ or $\alpha=\pm2s,ns$ $\beta=\pm$,
\beq
\Theta_{\alpha\beta}(\lambda_j^\alpha-\lambda_k^\beta)=
-i f_\alpha\ln\Bigl[\epsilon_{\alpha\beta}h(\lambda_j^\alpha-\lambda_k^\beta)
                                 h(\lambda_j^\alpha-\lambda_k^{\beta*})\Bigr]
\eeq
where $\epsilon_{+,-2s}=\epsilon_{-,2s}=-1$, $=1$ otherwise.
For $\alpha,\beta=\pm 2s,ns$
\beq
\Theta_{\alpha\beta}(\lambda_j^\alpha-\lambda_k^\beta)=
-i f_\alpha\ln\Bigl[\epsilon_{\alpha\beta}
                    h(\lambda_j^\alpha-\lambda_k^\beta)
                    h(\lambda_j^\alpha-\lambda_k^{\beta*})
                    h(\lambda_j^{\alpha*}-\lambda_k^\beta)
                    h(\lambda_j^{\alpha*}-\lambda_k^{\beta*})\Bigr]
\eeq
where $\epsilon_{2s,2s}=\epsilon_{-2s,-2s}=-1$, $1$ otherwise.  (Note
that in~\cite{adm}, the functions $t_\alpha$ and
$\Theta_{\alpha\beta}$ for $\alpha=\pm2s,ns$ were defined without the
factor $f_\alpha$.  This will change the completeness rules somewhat
from those presented in~\cite{adm}, but is necessary in order to have
positive densities.)  Here, all logarithms are taken so that $|{\rm
Im}\ \ln z|\leq \pi$, and the functions $t_\alpha$ and
$\Theta_{\alpha\beta}$ are defined so that
\beqa
t_\alpha(\lambda_j^\alpha) &=&
0 \qquad \hbox{if}\ \ \ {\rm Re}\ (\lambda_j^\alpha)=0,\nonumber\\
\Theta_{\alpha\beta}(\lambda_j^\alpha-\lambda_k^\beta)& =&
 0 \qquad \hbox{if}\ \ \
{\rm Re}\ (\lambda_j^\alpha)={\rm Re}\ (\lambda_k^\beta).
\eeqa

The logarithmic Bethe equations are written in terms of these functions:
\beq
Z(\lambda_j)\equiv \frac{I_j^\alpha}{M}=
\frac{1}{2\pi}t_\alpha(\lambda_j^\alpha)-
\frac{1}{2\pi M}\sum_{\beta=\pm,\pm2s,ns}\sum_{k=1}^{m_\beta}
\Theta_{\alpha\beta}(\lambda_j^\alpha-\lambda_k^\beta),
\label{logbae}
\eeq
where $I_j^\alpha$ are (half-) integers. We now present the
completeness rules for the integers $I_j^\alpha$.  It is only
necessary at this point to discuss the completeness rules for $Q=0$.
It will be shown in section 5 that the $Q=\pm1$ sectors are identical
to this sector in the thermodynamic limit.  The completeness rules
of~\cite{adm} for $Q=0$ in the notation introduced here become:
\begin{enumerate}
\item $I_k^+$ and $I_k^{2s}$ are distinct (half-)integers, are chosen from
the same set of $m_++m_{2s}$ integers, and $I_k^+=I_k^{2s\ h}$,
where $h$ represent a ``hole'' or missing integer. Therefore the set
$\{I_j^+\}+\{I_j^{2s}\}$ fills the interval
$-1/2(m_++m_{2s})$ to $1/2(m_++m_{2s})$.
\item $I_k^-$ and $I_k^{-2s}$ are distinct (half-)integers, are chosen from
the same set of $m_-+m_{-2s}$ integers, and $I_k^+=I_k^{2s\ h}$.
Again, the set $\{I_j^-\} + \{I_j^{-2s}\}$ fills the interval.
\item $I_k^{ns}$ are distinct (half-)integers chosen from a set
of $2m_-+2m_{-2s}+m_{ns}$ (half-)integers.
\item The spacing between ``available'' integers, the set of integers
$\{I_k^\alpha\}+\{I_k^{\alpha\ h}\}$, is 1.
\end{enumerate}
We see that $+$  integers correspond to missing $2s$ integers,
and the same for $-$ and $-2s$.  In addition to these rules, there is
a sum rule for $m_\alpha$:
\beqa
m_+=2n_{ns}+3m_-+4m_{-2s} \nonumber \\
m_{ns}+2m_{ns}+3m_{-2s}+2m_-=M
\label{particle numbers}
\eeqa
This sum rule is responsible for restricting the maximum integers
$I_{max}^\alpha$ as a function of the number of excitations in the
system.

We make the assumption at large $M$ that the rules (1) and (2) imply
the equality
\beq
\lambda_j^+ =\lambda_j^{2s\ h},\qquad \lambda_j^-=\lambda_j^{-2s\ h}.
\label{ass}
\eeq
This appears to be true from numerical results, has been proven for
order one excitations~\cite{adm2}, and shown to be consistent for
all excitation densities in the thermodynamic limit.

We now take the thermodynamic limit $M\rightarrow\infty$ of
the Bethe equations~(\ref{logbae}), with $\lambda$ fixed.  When we do
this, we lose the information contained in the rules (1)-(3) about the
maximum integers.  We rewrite the functions $\Theta_{\alpha,\beta}$ and
$t_\alpha$ in terms of the real part $\lambda_j^\alpha$,
using~(\ref{roottypes}), and take the derivative of $Z(\lambda)$ with
respect to $\lambda$ in the thermodynamic limit.  We obtain the
following set of equations:
\beqa
\rho_t^\pm(\lambda) &=& \frac{1}{\pi}K_{\pi/12}^\pm(\lambda) -
\frac{1}{2\pi} \Bigl[
K_{\pi/3}^\pm*(\rho_p^+ - \rho_p^{-2s}) +K_{\pi/3}^\mp*(\rho_p^--\rho_p^{2s})+
\{K_{\pi/12}^++K_{\pi/12}^-\}*\rho_p^{ns}\Bigr] \nonumber \\
\rho_t^{\pm2s}(\lambda) &=& \frac{1}{\pi}(K_{\pi/12}^\pm(\lambda)-
K_{\pi/4}^{\pm}(\lambda)) -
\frac{1}{2\pi} \Bigl[
K_{\pi/3}^\mp*(\rho_p^+ - \rho_p^{-2s}) +K_{\pi/3}^\pm*(\rho_p^--\rho_p^{2s})
\nonumber\\& & \hbox{\hskip1in} +
\{K_{\pi/12}^++K_{\pi/12}^-\}*\rho_p^{ns}\Bigr] \nonumber \\
\rho_t^{ns}(\lambda)&=&
-\frac{1}{2\pi}(K_{\pi/3}^+(\lambda)+K_{\pi/3}^-(\lambda))
+\frac{1}{2\pi}\Bigl[
\frac{1}{2}(K_{\pi/12}^++K_{\pi/12}^-)*(\rho_p^+-\rho_p^{-2s}+\rho_p^-
-\rho_p^{2s}) \nonumber\\
& &\hbox{\hskip1in} + (K_{\pi/3}^++K_{\pi/3}^-)*\rho_p^{ns}\Bigr]
\label{lbae}
\eeqa
where
\beq
\rho_t^\alpha=\lim_{M\rightarrow\infty}\frac{1}
{M(\lambda_{I_j+1}-\lambda_{I_j})},\qquad
\rho_p^\alpha =\lim_{M\rightarrow\infty}\frac{1}
{M(\lambda_{I_{j+1}}-\lambda_{I_j})},
\label{dendef}
\eeq
the convolution $*$ is defined as:
\beq
f*g = \int_{-\infty}^{\infty}d\mu f(\lambda-\mu)g(\mu),
\eeq
and the kernels $K_\alpha^\pm(\lambda)$ are
\beq
K_\alpha^\pm(\lambda)=
 \frac{\pm 2 \sin{2\alpha}}{\cosh{2\lambda}\mp\cos{2\alpha}}.
\eeq

In writing equations~(\ref{lbae}), we did not make use of the relationship
between holes and particles~(\ref{ass}), which imposes a relationship
between the densities in equations~(\ref{lbae}).  The
assumption~(\ref{ass}) implies that when particle integers of, say,
$+$ are equal to the ``hole'' integers of $2s$, their corresponding
rapidities are are equal.  Therefore, in light of the
definitions~(\ref{dendef}), the total densities of $+$ and $2s$ are
equal (and those of $-$ and $-2s$ as well), and the particle densities
are related in a simple way:
\beq
\rho_t^+(\lambda)=\rho_t^{2s}(\lambda), \quad
\rho_t^-(\lambda)=\rho_t^{-2s}(\lambda),\quad
\rho_p^{\pm2s}(\lambda)=\rho_t^\pm(\lambda)-\rho_p^\pm(\lambda)
\label{den1}
\eeq
This allows us to rewrite the density equations~(\ref{lbae}), in terms
of three independent particle densities.  It is convenient for further
computation do this separately for the ferromagnetic and
anti-ferromagnetic spin chains.

\section{Ferromagnetic chain}
\setcounter{equation}{0}
For the ferromagnetic hamiltonian, we know from~\cite{adm,adm2} that
the order one excitations are $+,-$ and $ns$.  We therefore choose to
rewrite~(\ref{lbae}) using~(\ref{den1}), as
\beqa
\rho_t^+(\lambda)&=& \frac{6}{\pi\cosh{6\lambda}} +
K_1 *(\rho_p^++\rho_p^-)- K_2*\rho_p^{ns},\nonumber\\
\rho_t^-(\lambda)&=&
K_1 *(\rho_p^++\rho_p^-)- K_2*\rho_p^{ns},\nonumber\\
\rho_t^{ns}(\lambda)&=& K_2 *(\rho_p^++\rho_p^-),
\label{bae+}
\eeqa
where the kernels are
\beq
K_1(\lambda) = \frac{18}{\pi^2}\frac{\lambda}{\sinh{6\lambda}},\qquad
K_2(\lambda) = \frac{3}{\pi\cosh{6\lambda}}.
\eeq
The particle densities in equations~(\ref{bae+}) above are now all independent
of each other, there are no additional constraints.

In the thermodynamic limit, the sum rule~(\ref{particle numbers})
becomes a relationship between total particle densities $D_\alpha$,
\beq
D_\alpha = \lim_{M\rightarrow\infty}\frac{m_\alpha}{M} =
\int d\lambda \rho_p^\alpha.
\eeq
However, we find we do not need to impose the sum rule as an
additional restriction on the densities in~(\ref{bae+}), as it is
contained in those equations already.  To see this, we take the
fourier transform of the first two equations in~(\ref{bae+}) and
evaluate at $k=0$. This gives exactly the relationship~(\ref{particle
numbers}) divided by $M$.

In~\cite{km}, the sum rules~(\ref{particle numbers}) were found to
give rise to the infrared anomaly, that is, to the diminishing of the
number of states near $P=0$ for the anti-ferromagnetic case, and thus
to exclusion rules beyond those of fermions.  Here, although we lose
information about how the maximum integers change as a function of
$m_\alpha$ when we take the thermodynamic limit, we still retain a
restriction between the densities which contains some of this
information.  This restriction will allow us to retain the concept of
correct counting of states in the thermodynamic limit.

The free energy is
\beq F = E-TS \label{fe} \eeq
evaluated at the stationary point with respect to independent particle
densities, where $S$ is the entropy of a state with fixed densities
$\rho_p^\alpha$ and $E$ is the total energy of the state.  for large $M$, the
entropy is:
\beq
S = M \sum_{\alpha=+,-,ns} \int_{-\infty}^{\infty} d\lambda
\Bigl(\rho_t^\alpha\ln{\rho_t^\alpha} -
\rho_p^\alpha\ln{\rho_p^\alpha} -
\rho_h^\alpha\ln{\rho_h^\alpha}
\Bigr),
\label{ent}
\eeq
where $\rho_h=\rho_t-\rho_p$.
The energy $E$ is the thermodynamic limit of equation~(\ref{energy}):
\beq
E = \sum_{\alpha=+,-,2s,-2s,ns} \int d\lambda \rho_p^\alpha(\lambda)
 e^\alpha(\lambda) -\frac{2 M}{\sqrt{3}}
\label{en1}
\eeq
where $e^\alpha(\lambda)$ is the energy associated with a root of type
$\beta$:
\beqa
e^\pm(\lambda) &=& \frac{\pm1-2i\sinh2\lambda}{2\cosh2\lambda\mp\sqrt3},
\qquad
e^{ns}(\lambda)=\frac{-2\sqrt3-4i\sinh4\lambda}{1+2\cosh4\lambda},\nonumber\\
& &e^{\pm2s}(\lambda) =\frac{\mp1-2i\sinh2\lambda}{2\cosh2\lambda\mp\sqrt3}
+\frac{\pm1-i\sinh2\lambda}{\cosh2\lambda}.
\eeqa
The energy in~(\ref{en1}) is not manifestly real.  However,
using~(\ref{bae+}), we find that the energy can be re-expressed in
terms of only the independent particle densities, and depends only on
$\rho_p^+$:
\beq
E=-\frac{2M}{\sqrt3} + \int_{-\infty}^{\infty} d\lambda
\frac{6}{\cosh6\lambda}\rho_p^+(\lambda).
\label{en+}
\eeq
This expression for the energy is manifestly real.
Minimizing~(\ref{fe}) with respect to the three particle densities
$\rho_p^+,\rho_p^-,\rho_p^{ns}$, we obtain the free energy per site,
\beq
\label{free energy}
f= e_0 - T \int_{-\infty}^{\infty}d\lambda \frac{6}{\pi\cosh{6\lambda}}
\ln\bigl(1+e^{-\epsilon^+(\lambda)/T}\bigr),
\eeq
where
\beq
e_0 = -\frac{2}{\sqrt3} + \int_{-\infty}^{\infty}d\lambda
\frac{6e^{2s}(\lambda)}{\pi\cosh6\lambda} =
-\frac{4}{\pi}-\frac{8}{3\sqrt3}=-2.81284\ldots
\eeq
and the densities $\epsilon^\beta$ are defined as
$\rho_h^\beta/\rho_p^\beta=\exp{(\epsilon^\beta/T)}$, and satisfy the
nonlinear integral equations:
\beqa
\epsilon^{+}(\lambda)&=&\frac{6}{\cosh6\lambda} -
T\Bigl[K_1 * \ln\Bigl[(1+e^{-\epsilon^+/T})(1+e^{-\epsilon^-/T})\Bigr]+
K_2 *\ln(1+e^{-\epsilon^{ns}/T})\Bigr],\nonumber \\
\epsilon^-(\lambda)&=&\epsilon^+(\lambda)-\frac{6}{\cosh6\lambda},\nonumber\\
\epsilon^{ns}(\lambda)&=&
T K_2 *\Bigl[\ln(1+e^{-\epsilon^+(\mu)/T})(1+e^{-\epsilon^-(\mu)/T})\Bigr].
\label{nie}
\eeqa
The functions $\epsilon^\alpha$ are also referred to as dressed
energies.  Note that~(\ref{nie}) represents only one integral equation
for $\epsilon^+$, since $\epsilon^-$ is simply related to
$\epsilon^+$, and the equation for $\epsilon^{ns}$ is not an integral
equation, as $\epsilon^{ns}$ does not appear on the right hand
side.

At fixed $\lambda$ and $T=0$, we get from~(\ref{nie})
\beq
\epsilon_0^+(\lambda)=\frac{6}{\cosh6\lambda},\qquad
\epsilon_0^-=\epsilon_0^{ns}=0,
\eeq
which are the order one excitations found in~\cite{adm2} for the
ferromagnetic chain.  The free energy per site in this limit is
$f=e_0$, which is the ground state energy found
in~\cite{baxter10,auyang,albert}.

The linear term in the specific heat $C$ is obtained from the low
temperature expansion of the free energy, or the entropy~(\ref{ent}):
\beq
\label{sph1}
C=-T\frac{\partial^2F}{\partial T^2}=T\frac{\partial S}{\partial T}.
\eeq
We find the linear term in the specific heat by computing the $O(T)$
term in the low temperature
entropy~\cite{filyov,babujian,kirillov,baz-resh}.  As $T\rightarrow0$,
$\epsilon^+(\lambda)/T$ scales as $1/(T \cosh 6\lambda)$, which vanishes
and gives no contribution to the integral~(\ref{free energy}) except
when $\lambda\sim O(\frac{1}{6}\ln T)$.  We rescale the
equations~(\ref{nie}) by making the change of variables
$\lambda\rightarrow\lambda-\frac{1}{6}\ln T$ and consider the
equations~(\ref{nie}) at large $\lambda$ and small $T$.  We define, at
this range of variables,
$\phi^\beta(\lambda)=\epsilon^\beta(\lambda-\frac{1}{6}\ln T)/T.$ The
integral equations become
\beqa
\phi^+(\lambda)&\simeq& 12 e^{-6\lambda} -
K_1*\ln\Bigl[(1+e^{-\phi^+})(1+e^{-\phi^-})\Bigr] - K_2*\ln(1+e^{-\phi^{ns}}),
\nonumber\\
\phi^-(\lambda)&\simeq&
-K_1*\ln\Bigl[(1+e^{-\phi^+})(1+e^{-\phi^-})\Bigr] - K_2*\ln(1+e^{-\phi^{ns}}),
\nonumber\\
\phi^{ns}(\lambda)&\simeq&
K_2*\ln\Bigl[(1+e^{-\phi^+})(1+e^{-\phi^-})\Bigr].
\label{phinie}
\eeqa
Differentiating~(\ref{phinie}) with respect to $\lambda$,
\beqa
\frac{d\phi^+}{d\lambda}&\simeq& -2\times 36 e^{-6\lambda} +
K_1* \Bigl[ \frac{\phi'^+}{1+e^{\phi^+}} +\frac{\phi'^-}{1+e^{\phi^-}}\Bigr]
+K_2 *\frac{\phi'^{ns}}{1+e^{\phi^{ns}}}\nonumber\\
\frac{d\phi^-}{d\lambda}&\simeq&
K_1* \Bigl[ \frac{\phi'^+}{1+e^{\phi^+}} +\frac{\phi'^-}{1+e^{\phi^-}}\Bigr]
+K_2 *\frac{\phi'^{ns}}{1+e^{\phi^{ns}}}\nonumber\\
\frac{d\phi^{ns}}{d\lambda}&\simeq&
-K_2* \Bigl[ \frac{\phi'^+}{1+e^{\phi^+}} +\frac{\phi'^-}{1+e^{\phi^-}}\Bigr]
\label{dphi}
\eeqa
where $\phi'^\beta=d\phi^\beta/d\lambda$.

We rescale equations~(\ref{bae+}) in the same way.  Let
$\tilde\rho_t^\beta(\lambda)=\rho_t^\beta(\lambda-\frac{1}{6}\ln T)$,
and recall that $\rho_p^\beta=\rho_t^\beta/(1+e^{\epsilon^\beta/T})$.  Then
\beqa
\tilde\rho_t^+(\lambda)&=& \frac{12 T}{\pi}e^{-6\lambda} +
K_1*(\frac{\tilde\rho_t^+}{1+e^{\phi^+}}
+\frac{\tilde\rho_t^-}{1+e^{\phi^-}})
-K_2*\frac{\tilde\rho_t^{ns}}{1+e^{\phi^{ns}}}\nonumber\\
\tilde\rho_t^-(\lambda)& = & K_1*(\tilde\frac{\rho_t^+}{1+e^{\phi^+}}
+\frac{\tilde\rho_t^-}{1+e^{\phi^-}})
- K_2*\frac{\tilde\rho_t^{ns}}{1+e^{\phi^{ns}}}\nonumber\\
\tilde\rho_t^{ns}(\lambda)&=&
K_2*(\frac{\tilde\rho_t^+}{1+e^{\phi^+}}+\frac{\tilde\rho_t^-}{1+e^{\phi^-}}).
\label{bae+r}
\eeqa
Comparing equation~(\ref{bae+r}) to~(\ref{dphi}),  we see that
\beq
\tilde\rho_p^\pm=-\frac{T}{6\pi}
\frac{d\phi^\pm}{d\lambda}
\frac{1}{1+e^{\phi^\pm}},\qquad
\tilde\rho_p^{ns}=\frac{T}{6\pi}
\frac{d\phi^{ns}}{d\lambda}
\frac{1}{1+e^{\phi^{ns}}}.
\label{rhoasym}
\eeq

The entropy can be evaluated in this limit. The
$\lambda\rightarrow\infty$ and  $\lambda\rightarrow-\infty$ limits
make the same contribution to $S$ (and $f$).  Therefore we write
\beqa
S&\simeq&2\sum_{\beta=+,-,ns}\inte\lambda \Bigl\{
\tilde\rho_p^\beta\ln(1+e^{\phi^\beta})
+\tilde\rho_h^\beta\ln(1+e^{-\phi^\beta})\Bigr\}\nonumber\\
&=&2\times \frac{T}{6\pi}\Bigl[
-\int_{\phi^+(-\infty)}^{\phi^+(\infty)}d\phi g(\phi)-
\int_{\phi^-(-\infty)}^{\phi^-(\infty)}d\phi g(\phi) +
\int_{\phi^{ns}(-\infty)}^{\phi^{ns}(\infty)}d\phi g(\phi) \Bigr],
\label{asyment}
\eeqa
where
\beq
g(\phi)= \frac{\ln(1+e^\phi)}{1+e^\phi}+\frac{\ln(1+e^{-\phi})}{1+e^{-\phi}}
\label{g}
\eeq
The limits $\phi(\pm\infty)$ are found from equations~(\ref{phinie}).
In these limits, the integrals can be performed by taking the log out
from under the integral sign, and integrating only the kernel.  Let
$\tilde\phi$ denote the asymptotic value under consideration
$\phi(\lambda=\pm\infty)$.  Then, for $\lambda=
\infty$, we obtain a system of equations
\beqa
\tilde\phi^+ =\tilde\phi^-&=&
-\frac{1}{4}\ln[(1+e^{-\tilde\phi^+})(1+e^{-\tilde\phi^-})]
-\frac{1}{2}\ln(1+e^{-\tilde\phi^{ns}}),\nonumber\\
\tilde\phi^{ns} &=& \frac{1}{2}\ln[(1+e^{-\tilde\phi^+})(1+e^{-\tilde\phi^-})].
\eeqa
Therefore the upper limits of equation~(\ref{asyment}) are
\beq
\phi^+(\infty)=\phi^-(\infty)=-\ln2,\qquad\phi^{ns}(\infty)=\ln3.
\eeq
At $\lambda=-\infty$, $\tilde\phi^+=\infty$, and
\beqa
\tilde\phi^- &=& -\frac{1}{4}\ln(1+e^{-\tilde\phi^-})
 -\frac{1}{2}\ln(1+e^{-\tilde\phi^{ns}}),\nonumber\\
\tilde\phi^{ns}&=& \frac{1}{2}\ln(1+e^{-\tilde\phi^-}),
\eeqa
and thus the lower limits in~(\ref{asyment}) are
\beq
\phi^+(-\infty) =\infty,\qquad
\phi^-(-\infty) =-\ln\Bigl[\frac{1+\sqrt5}{2}\Bigr],\qquad
\phi^{ns}(-\infty) =\ln\Bigl[\frac{1+\sqrt5}{2}\Bigr]
\eeq

We show in the appendix that how to express the
integral~(\ref{asyment})  in terms of dilogarithms.
Using the identities described in the appendix, we find that in the low
temperature limit,
\beq
S\simeq \frac{4\pi T}{45},\qquad
\eeq
and from~(\ref{spheat}) we see that, with $v_F=3$~\cite{adm2}, the
central charge $c=\frac{4}{5}$.  This is the central charge of the
well known conformal limit of the 3-state Potts
chain~\cite{huse,dotsenko,bpz} computed in the limit~(\ref{cft limit}).
This verifies that the limits~(\ref{cft limit}) and~(\ref{tba limit})
smoothly connect together, and there are no additional length scales in
the problem.

In the calculation above, we find that, although in the zero
temperature limit the energies $\epsilon^-,\epsilon^{ns}$ vanish, they
contribute to the low temperature specific heat, i.e. the functions
$\phi^{-}(\lambda),\phi^{ns}(\lambda)$ do not vanish.  This is a
manifestation of the feature seen in~\cite{adm2}, that although the
energy can be expressed, as in~(\ref{en+}), in terms of only
$\rho_p^+$, the number of states $e^S$ with energy $E$ depends on
$\rho_t^\alpha$, and therefore on $\rho_p^-$ and $\rho_p^{ns}$, as it
depends on $m_-,m_{ns}$ in~\cite{adm2}.  In our case, the densities
$\rho_p^-,\rho^{ns}$ enter the expression for the free energy via the
entropy $S$, which counts the states.  Computation of the specific
heat at low temperature depends sensitively on correct counting of
states, as was seen in~\cite{km} for the anti-ferromagnetic case.
Note that from the counting rules of~\cite{adm}, and from the
equations~(\ref{bae+}), the number of states near $P=0$ increases as
the number of excitations increase.  This causes the central charge to
be larger than 1/2, the value we would expect if the excitations
$\rho_p^+$ were fermions.

\section{Anti-ferromagnetic chain}
\setcounter{equation}{0}
To find the low temperature behavior of the anti-ferromagnetic chain,
we rewrite~(\ref{lbae}) in term of the $\pm2s,ns$ densities, which we
know to be the order one excitations for this
hamiltonian~\cite{adm,adm2,km}. The equations~(\ref{lbae}) become
\beqa
\rho_t^{\pm2s}(\lambda) &=& \frac{3}{\pi}\frac{1}{\sqrt2\cosh3\lambda\mp1}
-\overline{K}_1*(\rho_p^{2s}+\rho_p^{-2s})-
\overline{K}_2*\rho_p^{ns},\nonumber\\
\rho_t^{ns}(\lambda)&=&\frac{3}{\pi\cosh3\lambda} -
\overline{K}_2*(\rho_p^{2s}+\rho_p^{-2s})-2\overline{K}_1*\rho_p^{ns},
\label{bae-}
\eeqa
where the kernels are:
\beq
\overline{K}_1(\lambda)=\frac{3}{2\pi\cosh3\lambda},\qquad
\overline{K}_2(\lambda)=\frac{6\cosh3\lambda}{\sqrt2\pi\cosh6\lambda}.
\label{kernels2}
\eeq
Note that, from these equations, in the antiferromagnetic
case the density of available states always diminishes with increasing
particle densities.  This was seen in~\cite{adm,km}, where the number
of available states decreases as a function of $m_{2s},m_{-2s},m_{ns}$.

The entropy in terms of these densities looks the same as~(\ref{ent}),
but now we sum over the three independent densities $\alpha=\pm2s,ns$.
We also express $E$ in terms of the $\pm2s,ns$ densities.  Minimizing
the quantity~(\ref{fe}) with respect to the three particle densities
$\rho_p^{2s},\rho^{-2s},\rho^{ns}$ now gives the free energy in terms
of these densities:
\beq
f=\tilde{e}_0+
T\inte\lambda\Bigl\{
\frac{3\ln(1+e^{-\epsilon^{2s}(\lambda)/T})}{\pi(\sqrt2\cosh3\lambda-1)}
+\frac{3\ln(1+e^{-\epsilon^{-2s}(\lambda)/T})}{\pi(\sqrt2\cosh3\lambda+1)}
+\frac{3\ln(1+e^{-\epsilon^{ns}(\lambda)/T})}{\pi\cosh3\lambda}\Bigr\},
\label{free energy-}
\eeq
where
\beq
\tilde{e}_0=e_0+\frac{18}{\pi^2}\inte\lambda\frac{1}{\cosh6\lambda(
\sqrt2\cosh3\lambda-1)} = 3-\frac{8}{3\sqrt3}+\frac{2}{\pi} =2.097\ldots,
\eeq
and $\epsilon^\alpha$ satisfy the integral equations
\beqa
\epsilon^{\pm2s}(\lambda)&=& \frac{3}{\sqrt2\cosh3\lambda\mp1} +
T\Bigl\{\overline{K}_1*\ln\bigl[(1+e^{-\epsilon^{2s}/T})(1+e^{-\epsilon^{-2s}/T})]
+\overline{K}_2*\ln(1+e^{-\epsilon^{ns}/T})
\Bigr\},\nonumber \\
\epsilon^{ns}(\lambda) &=& \frac{3}{\cosh3\lambda} + T\Bigl\{
\overline{K}_2*\ln\Bigl[(1+e^{-\epsilon^{2s}/T})(1+e^{-\epsilon^{-2s}/T})]
+2\overline{K}_1*\ln(1+e^{-\epsilon^{ns}/T})
\Bigr\}.
\label{nie-}
\eeqa

At $T=0$, we see from~(\ref{nie-}) that
\beq
\epsilon_0^{\pm2s} = \frac{3}{\sqrt2\cosh3\lambda\mp1},\qquad
\epsilon_0^{ns}    = \frac{3}{\cosh3\lambda},
\label{ooe-}
\eeq
and from~(\ref{free energy-}) the free energy is $f=\tilde e_0$.
These are the order one excitations and ground state energy for the
anti-ferromagnetic hamiltonian found in~\cite{baxter10,adm2}.

Again we compute the low temperature limit of the entropy, but now
 we rescale the integral equations by changing variables to
$\lambda\rightarrow\lambda-\frac{1}{3}\ln T$.  Defining
$\phi^\beta(\lambda)=\epsilon(\lambda-\frac{1}{3}\ln T)/T$ we have
\beqa
\phi^{2s}=\phi^{-2s}&\simeq& 2\times\frac{3}{\sqrt2} e^{-3\lambda}
+ T\Bigl\{ K_1 *\ln\Bigl[(1+e^{-\phi^{2s}})(1+e^{-\phi^{-2s}})\Bigr]
+K_2 * \ln(1+e^{-\phi^{ns}})\Bigl\}\nonumber\\
\phi^{ns}&\simeq& 2\times3 e^{-3\lambda}
+ T\Bigl\{ K_2 *\ln\Bigl[(1+e^{-\phi^{2s}})(1+e^{-\phi^{-2s}})\Bigr]
+2K_1 * \ln(1+e^{-\phi^{ns}})\Bigl\},
\label{phi-}
\eeqa
so that $2s$ and $-2s$ are symmetric in this limit, which was a feature
seen in~\cite{km}.
Again, differentiating~(\ref{phi-}) with respect to $\lambda$ and comparing
to the density equations~(\ref{bae-}) rescaled as $\lambda\rightarrow\lambda
-\frac{1}{3}\ln T$, we see that
\beq
\tilde\rho_p^{\beta} = -\frac{T}{3\pi}\frac{d\phi^\beta}{d\lambda}
\frac{1}{1+e^{\phi^\beta}},\ \ \ \beta=2s,-2s,ns.
\eeq

The entropy is calculated as in~(\ref{asyment}).
{}From~(\ref{phi-}) we find the limits $\phi^\alpha(\pm\infty)$
\beqa
\phi^{2s}(-\infty)&=&\phi^{-2s}(-\infty)=\phi^{ns}(-\infty)=\infty,\nonumber\\
\phi^{-2s}(\infty)&=&\phi^{2s}(\infty) = \ln 2,\qquad  \phi^{ns}(\infty)=\ln 3.
\eeqa
The entropy is
\beq
S\simeq -\frac{2 T}{3\pi}\Bigl[ 2\int_{\infty}^{\ln2}g(\phi)d\phi +
\int_{\infty}^{\ln3}g(\phi)d\phi
\Bigr],
\eeq
which, using the dilogarithmic identities in the appendix, gives
\beq
S\simeq = \frac{2\pi T}{9}.
\label{ent-}
\eeq
This, with $v_F=3/2$~\cite{adm2}, gives a central charge  $c=1$,
which  is the central charge of the conformal limit of the
model~\cite{park-widom,baz-resh}. Again, this verifies that the
limits~(\ref{cft
limit}) and~(\ref{tba limit}) commute, and there are no additional
length scales.  We see that the fact that $\rho_t^\alpha$ tends to
decrease with increasing $\rho_p^\alpha$ causes the central charge to
be smaller than the value 3/2 one would expect for pure fermions.

\section{$\bf Q=\pm1$}

In the thermodynamic limit, quantities which are not of order $M$ are
irrelevant to the calculation.  Therefore, we do not expect the value
of $Q$ to affect the thermodynamic equations.  This is indeed the
case.

In~\cite{adm}, the counting rules for $Q=\pm1$ were found.  It was shown
that the counting rules depended on the value of the numbers $m_{++}$ and
$m_{-+}$, where
\beq
m_{++}-m_{-+} = 0, \pm1.
\label{diff}
\eeq
Again we note that this difference is not of order $M$ and we do not
expect it to change the thermodynamic equations.  The sum
rules~(\ref{particle numbers}) for this sector are changed to:
\beqa
m_+ &=& 2 m_{ns} + 4 m_{-2s} + m_{-+} + m_{++} \nonumber \\
M-1 &=& m_{2s} + 2 m_{ns} + 3 m_{-2s} + m_{-+} + m_{++}.
\eeqa
This is only different from the $Q=0$ sector by a term of order 1,
due to equation~(\ref{diff}) and the fact that $m_{-+}=m_-$.  Therefore
to order $M$, the sum rules are identical to~(\ref{particle numbers}).

For $m_{++}-m_{-+}=\pm1$, the completeness rules for $Q=1$ are the
same as for $Q=0$.  For $m_{++}-m_{-+}=0$, there is spectrum doubling:
The integers $I_j^{2s}$ are shifted from those of $+$ by $\pm1/2$, and
those of $-2s$ are shifted from $I_j^-$ in the same way, both signs
giving the same energy level.  The shift does not affect the
thermodynamic limit of equations~(\ref{logbae}).  The spectrum
doubling gives rise to an additive term of order one in the entropy
(which counts the number of states).  Since the entropy is of order
$M$, again this term is not relevant in the thermodynamic limit.  We
conclude therefore that this sector is identical to $Q=0$.

In~\cite{km}, the difference in counting rules for the sectors
$Q=\pm1$ gave rise to different branching functions in the modular
invariant partition function from those of $Q=0$.  However, each term
in the modular invariant partition function gives the same specific
heat, Due to the modular invariance property.  The specific heat is
found from the limit $q\rightarrow1$ of the partition function
of~\cite{km}.  However, the partition function is invariant under
modular transformations, where, if $q=\exp(2\pi i\tau)$, the
transformation $\tau\rightarrow-1/\tau$ leaves the partition function
invariant, so the specific heat is obtained from the $q\rightarrow0$
limit of the partition function. The same transformation sends each
branching function into a linear combination of all other branching
functions.  Therefore, each branching function has the same
$q\rightarrow1$ behavior. In the thermodynamic calculation we do not
see this difference between the sectors $Q=0$ and $Q=\pm1$.

\section{Discussion}
\setcounter{equation}{0}
In~\cite{km}, the order one excitations~(\ref{ooe-}) were used to
compute the partition function of the anti-ferromagnetic chain in the
limit~(\ref{cft limit}). This is identical to the modular invariant
partition function of the conformal limit of the
model~\cite{pearceun,gq} and so gives the same specific heat
as~(\ref{ent-}).  For the computation in~\cite{km}, it was necessary
to carefully consider the information about the way the maximum
integer $I_{max}^\alpha$ of the finite size system in~(\ref{logbae})
changed with the number of particles in the system.  This is because
those integers correspond to the energies close to zero, $e(P)\sim0$,
or in the language used here, the large $\lambda$ behavior of the
energies, which is the region which contributes to the specific heat.
The maximum integers decrease, at finite $M$, as more particles are
added to the system.  In~\cite{km} this phenomenon was referred to as
an {\em infrared anomaly}.  This represents counting rules for the
excitations beyond the fermionic exclusion rule, and is the phenomenon
responsible for the central charge being different from that of
fermions. In the case of the anti-ferromagnetic chain discussed
in~\cite{km}, the infrared anomaly was repulsive: fewer states were
available as the number of excitations was increased than would be
available for fermions.  In the case of the ferromagnetic chain, both
repulsive and attractive infrared anomalies are present, but the total
infrared anomaly is attractive.

In the thermodynamic limit, we discard the information about the
maximum integers $I_j^\alpha$.  Nevertheless, the integral
equations~(\ref{nie-}) contain the information about the way the
density of available states depends on the particle density,
represented by the density equations~(\ref{bae-}).  This information
enables correct counting of states, using the entropy.  We see that
this information gives the same specific heat as the counting of
ref~\cite{km}, but, as we saw in section 5, does not show the
difference between the different $Q$ sectors.

When computing the low temperature specific heat, the region of
$\lambda$ which contributes to the free energy as $T\rightarrow0$ is
the $\lambda\sim \frac{1}{6}\ln T$ limit in the ferromagnetic case,
and the $\lambda\sim\frac{1}{3}\ln T$ limit in the anti-ferromagnetic
case.  This limit is the $P\rightarrow0$ limit~\cite{adm2}, which
corresponds to the lowest lying order one excitations in~\cite{km}.
The careful counting of states there is paralleled in the computation
here by the rescaling of the integral equations and the density
equations in the limit $T\rightarrow0$ and $\lambda\rightarrow\infty$.

Finally, we note that these computations are related to the
thermodynamic Bethe ansatz method of~\cite{zamolodchikov}, a point
which is discussed in some detail in~\cite{kkmm}.

\section*{Acknowledgements}
We are pleased to acknowledge useful discussion with Dr. G. Albertini,
Dr. S. Dasmahapatra, and E. D. Williams.  We are particularly grateful
to  Prof. B. M. McCoy for his many suggestions and insights. We'd also
like to thank Prof. P. A. Pearce and Prof. V. V. Bazhanov for their comments.
This work was
partially supported by the National Science Foundation under grant
DMR-9106648.

\begin{appendix}
\section{Appendix: Expression of entropy via dilogarithms}
\setcounter{equation}{0}
The Rogers dilogarithm is defined as~\cite{rogers}
\beq
L(x) = -\frac{1}{2} \int_0^x df \left( \frac{\ln(1-f)}{f} + \frac{\ln(f)}{1-f}
\right).
\eeq
Making a change of variables in the expression for the
entropy~(\ref{asyment}) to $f=1/1+e^\phi$, the entropy~(\ref{asyment})
is expressed in terms of $L(x)$~\cite{baz-resh}:
\beq
S\simeq -\frac{2 T}{3 \pi} \left(
2 L(\frac{1}{3}) + L(\frac{1}{4}) - L(1) - 2 L(\frac{3-\sqrt 5}{2})\right).
\label{id1}
\eeq
We use the identity on Rogers dilogarithms~\cite{kirillov}:
\beq
\sum_{k=2}^{n-2} L\left(\frac{\sin^2(\pi/n)}{\sin^2(k\pi/n)}\right) =
\frac{2(n-3)}{n} L(1),
\eeq
where $L(1) = \pi^2/6$~\cite{lewin}. Using this identity with, $n=6$,
we find that
\beq
2 L(1/3) + L(1/4) = L(1),
\label{id2}
\eeq
and, with $n=5$, we see that~\cite{lewin}:
\beq
2 L\left(\frac{3-\sqrt5}{2}\right) = 2 \frac{\pi^2}{15}.
\eeq
Therefore equation~(\ref{id1}) gives
\beq
S\simeq \frac{4 T}{3\pi} L\left(\frac{3-\sqrt 5}{2}\right) = \frac{4 \pi T}
{45}.
\eeq

In the anti-ferromagnetic case, we again use~(\ref{id2}) with $n=6$,
and entropy is:
\beq
S\simeq \frac{4 T}{3\pi} (2 L(1/3) + L(1/4)) = \frac{4 T}{3\pi}L(1) =
\frac{2\pi T}{9}.
\eeq

\end{appendix}

\end{document}